\documentclass[fleqn,10pt]{wlscirep}
\usepackage[utf8]{inputenc}
\usepackage[T1]{fontenc}
\usepackage{subcaption}
\usepackage{multirow}
\usepackage{mathrsfs}
\usepackage{threeparttable}
\usepackage{array}
\usepackage{amsmath}
\usepackage{supertabular}
\usepackage{algorithm}
\usepackage{algorithmicx}
\usepackage{algpseudocode}
\usepackage{caption}
\usepackage{hyperref}
\usepackage[cal=cm]{mathalfa}
\usepackage{tabularx}
\usepackage{longtable}

\title{Continuous Telemonitoring of Heart Failure using Personalised Speech Dynamics}

\author[1,\#]{Yue Pan}
\author[2,3,*,\#]{Xingyao Wang}
\author[4,\#]{Hanyue Zhang}
\author[5,*]{Liwei Liu}
\author[5]{Changxin Li}
\author[2,6]{Gang Yang}
\author[5]{Rong Sheng}
\author[1]{Yili Xia}
\author[4,6,*]{Ming Chu}
\affil[1]{School of Information Science and Technology, Southeast University, Nanjing, China}
\affil[2]{Department of Cardiology, Chongqing Hospital of Jiangsu Province Hospital, Chongqing, China}
\affil[3]{Department of Cardiology, The First Affiliated Hospital of Chongqing Medical University, Chongqing, China}
\affil[4]{Department of Cardiology, The Affiliated Taizhou People's Hospital of Nanjing Medical University, Taizhou School of Clinical Medicine, Nanjing Medical University, Taizhou, China}
\affil[5]{Advanced Computing and Storage Laboratory, 2012 Laboratories, Huawei Technologies Co. Ltd.}
\affil[6]{Division of Cardiology, The First Affiliated Hospital of Nanjing Medical University, Nanjing, China}
\affil[$\#$]{These authors contributed equally.}
\affil[*]{Corresponding authors, e-mails: Ming Chu: \url{chuming@njmu.edu.cn}, Xingyao Wang: \url{mato00@163.com}, and Liwei Liu: \url{liuliwei5@huawei.com}}

\begin{abstract}
Remote monitoring of heart failure (HF) via speech signals provides a non-invasive and cost-effective solution for long-term patient management. However, substantial inter-individual heterogeneity in vocal characteristics often limits the accuracy of traditional cross-sectional classification models. To address this, we propose a Longitudinal Intra-Patient Tracking (LIPT) scheme designed to capture the trajectory of relative symptomatic changes within individuals. Central to this framework is a Personalised Sequential Encoder (PSE), which transforms longitudinal speech recordings into context-aware latent representations. By incorporating historical data at each timestamp, the PSE facilitates a holistic assessment of the clinical trajectory rather than modelling discrete visits independently. Experimental results from a cohort of 225 patients demonstrate that the LIPT paradigm significantly outperforms the classic cross-sectional approaches, achieving a recognition accuracy of 99.7\% for clinical status transitions. The model's high sensitivity was further corroborated by additional follow-up data, confirming its efficacy in predicting HF deterioration and its potential to secure patient safety in remote, home-based settings. Furthermore, this work addresses the gap in existing literature by providing a comprehensive analysis of different speech task designs and acoustic features. Taken together, the superior performance of the LIPT framework and PSE architecture validates their readiness for integration into long-term telemonitoring systems, offering a scalable solution for remote heart failure management.
\end{abstract}
\begin{document}

\flushbottom
\maketitle
\thispagestyle{empty}

\section{Introduction}
%背景
Heart failure (HF) is a progressive condition characterised by a decline in the heart's ability to pump blood effectively, affecting 26 million individuals worldwide \cite{amir2022remote}. In 2023, the number of Chinese HF patients hit 8.9 million, costing over 10,000 yuan per person for each hospitalised treatment \cite{Liu2025Interpretation}. Moreover, cardiovascular problems have a larger impact in areas with less abundant medical resources, like rural areas \cite{Liu2025Interpretation}. To mitigate rehospitalisation rates and address healthcare disparities in resource-limited regions, there is an urgent need for cost-effective, validated, and accessible methodologies for the continuous out-of-hospital assessment and monitoring of HF.

%语音与HF关联的医学成因与统计证实
Impaired cardiovascular function often leads to systemic fluid retention, including laryngeal oedema, which alters the mechanics of phonation. Although these acoustic variations are subtle in the early phase, they remain discernible through speech analysis \cite{murton2023acoustic}. Statistical evidence has demonstrated significant correlations between HF and various speech-derived features, including perturbation measures such as jitter, shimmer, and the harmonics-to-noise ratio \cite{murton2017acoustic}, as well as glottal source features reflecting altered vocal fold dynamics \cite{mittapalle2022glottal}. Furthermore, Schöbi et al. \cite{schobi2022evaluation} reported that speech fluency metrics are strongly associated with HF severity. 

%目前成果
Driven by the correlation between HF and vocal biomarkers, recent research has extensively validated the use of speech-based analysis for automated HF detection and risk stratification. To date, the majority of HF assessment frameworks rely on a conventional two-stage approach, combining manual feature engineering with supervised learning. While classical paradigms, such as support vector machines (SVM), extra trees (ET), and AdaBoost \cite{reddy2021automatic,windmon2018tussiswatch} achieved early advances, the field has increasingly pivoted toward deep learning to automate feature representation \cite{priyasad2022detecting,murton2023acoustic,firmino2023heart}. While these contributions establish the technical foundation for automated screening, their efficacy remains largely confined to controlled experimental settings. Crucially, such cross-sectional approaches \cite{liu2022inter} often fail to account for the subtle longitudinal fluctuations inherent in individual patient trajectories \cite{pan2025chinese}.

% 存在问题
Speech signals exhibit pronounced inter-individual variability, driven not only by intrinsic demographic determinants such as age, sex, and accent, but also by confounding comorbidities. These include respiratory impairments \cite{markova2016age, farrus2021speech, nallanthighal2022detection}, concomitant cardiovascular pathologies \cite{pareek2016coronary}, and cognitive-linguistic disorders \cite{toth2015automatic, gosztolya2016detecting}. Such extensive heterogeneity can mask the subtle vocal manifestations of HF, thereby substantially undermining the diagnostic robustness and cross-population generalisability of conventional classification models. This observation aligns with our preliminary findings \cite{pan2025chinese}.

% 提出解决方案
\begin{figure*}[]
\centering
\includegraphics[width=1.01\textwidth,trim=40 100 0 0,clip]{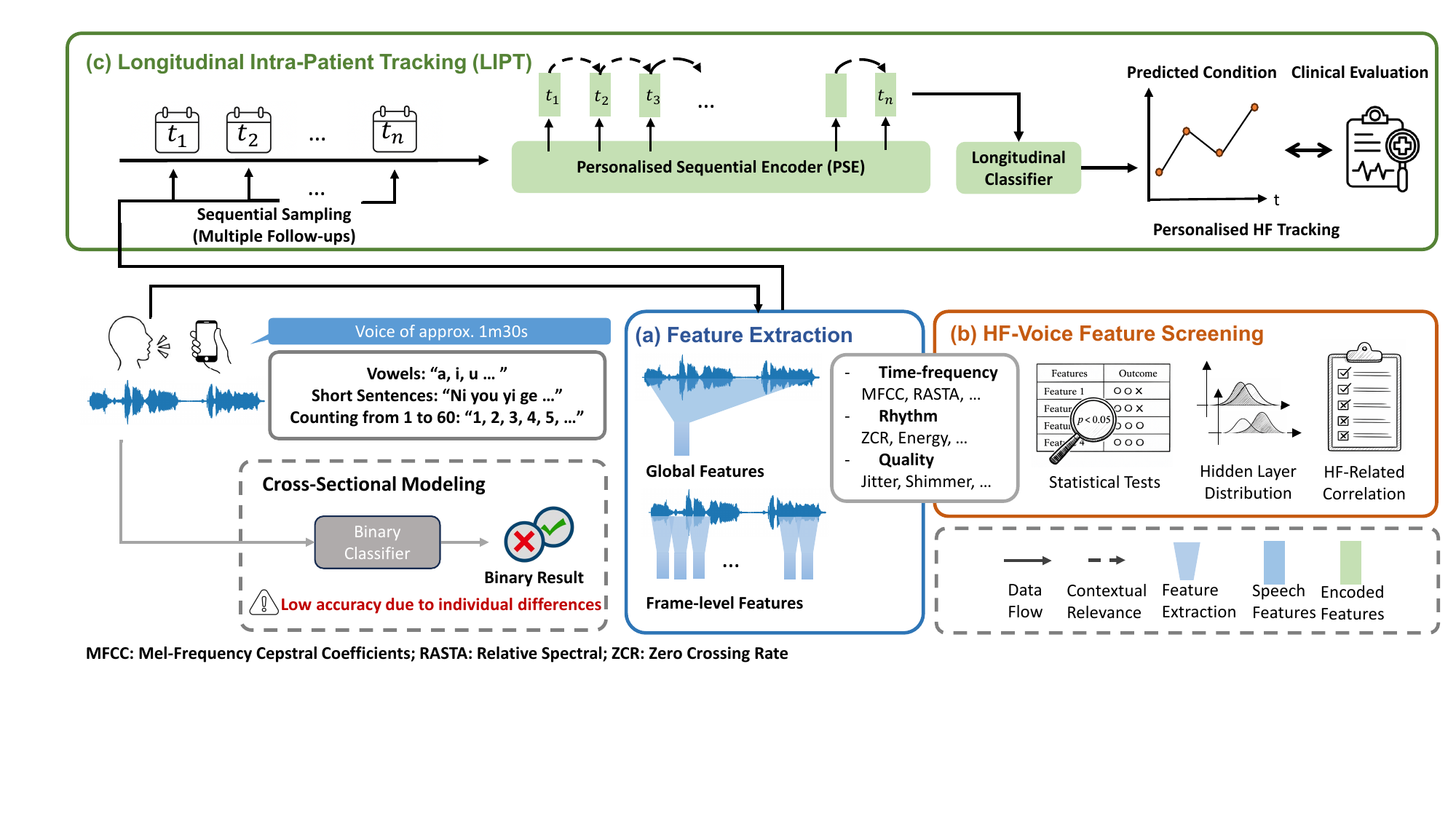}
\caption{Conceptual framework of the Longitudinal Intra-Patient Tracking (LIPT) paradigm for personalised HF monitoring. Unlike standard cross-sectional classification, which suffers from low accuracy due to inter-subject heterogeneity, the LIPT pipeline directly models intra-individual temporal sequences. The workflow comprises three primary stages: (a) Feature Extraction, in which speech signals are characterised by spectral (time-frequency), rhythmic, and glottal quality attributes at both global and frame-wise levels; (b) HF-Voice Feature Screening, which employs statistical significance tests to identify features with pathophysiological associations to HF, while correlations between various feature categories and the disease state are evaluated; and (c) Longitudinal Intra-Patient Tracking (LIPT), where sequential voice samples from each individual are processed by a Personalised Sequential Encoder (PSE) to capture longitudinal dependencies. Finally, a longitudinal classifier generates individualised tracking results, which are validated against clinical gold standards to monitor disease trajectories over time. }\label{motivation}
\end{figure*}

To overcome these challenges, we introduce a holistic framework for speech-based HF monitoring that prioritises individualised temporal trajectories over static, population-level benchmarks. As is shown in Figure \ref{motivation}, our contributions are threefold: First, addressing the absence of standardised protocols, we identify optimal configurations from both task-design and feature-selection perspectives to provide systematic guidance for the field. Second, to mitigate the confounding effects of inter-individual heterogeneity, we propose the Longitudinal Intra-Patient Tracking (LIPT) paradigm, which decouples HF-specific changes from inherent vocal characteristics. Finally, we develop a Personalised Sequential Encoder (PSE) to capture complex temporal dependencies across a patient’s historical speech data. By facilitating the modelling of longitudinal sequences of arbitrary length, the PSE ensures robust condition tracking without necessitating extrinsic population data. The efficacy of this integrated pipeline is rigorously validated through benchmarking and additional follow-up analysis, establishing a reliable pathway for long-term remote HF assessment.

\section{Results}

\subsection{Patient Enrollment}
\label{results_enrollment}

\begin{table}[h!]
\centering
\begin{threeparttable}
\caption{Speech tasks and their respective data quantity. A total of 7 tasks were attempted, including 3 vowels, 3 short sentences, and 1 long sentence.} 
\label{task_definition}
\setlength{\tabcolsep}{8pt} % 稍微加宽列间距，增强可读性
\begin{tabular}{ll l ccc}
\toprule
\multirow{2}{*}{\textbf{Task}} & \multirow{2}{*}{\textbf{Abbr.}} & \textbf{Chinese Pinyin} & \multicolumn{3}{c}{\textbf{Data Quantity}} \\
\cmidrule(lr){4-6}
& & \textbf{(Phonetic symbols)} & adm./dis. & f-up (stable) & f-up (re-ad.) \\ 
\midrule
Vowel 1       & a      & -                                     & 218 & 70 & 17 \\
Vowel 2       & u      & -                                     & 213 & 70 & 18 \\
Vowel 3       & i      & -                                     & 220 & 72 & 18 \\
\addlinespace
Short Sent. 1 & pg     & shan dong de ping guo you da you tian & 217 & 70 & 19 \\
Short Sent. 2 & mm     & ni you yi ge mei li de mei mei        & 197 & 67 & 16 \\
Short Sent. 3 & mlh    & hao yi duo mei li de mo li hua        & 191 & 65 & 17 \\
\addlinespace
Long Sent.    & count  & (numbers 1-60)                        & 213 & 71 & 18 \\ 
\bottomrule
\end{tabular}

\begin{tablenotes}[flushleft]
\footnotesize
\item[a] adm. = admission; dis. = discharge; f-up = follow-up; re-ad. = readmission; sent. = sentence.
\end{tablenotes}
\end{threeparttable}
\end{table}

This study utilised speech data from 225 patients recruited at Taizhou People’s Hospital (Jiangsu, China) who were hospitalised for acute decompensated heart failure (ADHF). Participants were required to perform the following vocal tasks: six sustained vowels, three short sentences, and one long sentence (counting from 1 to 60 in Mandarin). 

For the 225 patients included, data were collected across both decompensated and post-treatment states. Additionally, 78 patients who remained stable after being discharged from the hospital participated in a follow-up data collection, while 19 patients who were readmitted during the follow-up phase due to clinical deterioration also produced voice records. An overview of the speech tasks and their respective data quantities is presented in Table \ref{task_definition}. The comprehensive collection protocol and inclusion criteria are detailed in Section \ref{collection}. For each recording, 6,373 dimensions of global features and 72 dimensions of frame-level features were extracted. The technical specifications and grouping strategies for these features are provided in Section \ref{extraction}. The baseline methods \cite{pan2025chinese} (results in Section \ref{ind_pair_compare}) were designed to deploy global features only, while a combination of global and frame-level features is supported in the proposed PSE (structrual details see Section \ref{pse}, results in Sections \ref{task_comparison}, \ref{pse_result}, and \ref{followup_result_text}). 

For the experimental setup, 180 patients (80\%) were allocated to the training set, while the remaining 45 (20\%) constituted the test set. All subsequent follow-up records were reserved for additional evaluation. Throughout this study, a "positive class" denotes a deterioration in the patient’s HF condition, whereas a "negative class" indicates clinical improvement. Specifically, during the initial training and testing phases (Sections \ref{ind_pair_compare}, \ref{task_comparison}, and \ref{pse_result}), it is assumed that the HF condition improves from a decompensated state to a post-treatment state. In these instances, positive and negative samples are generated by the random allocation of input sequences (as detailed in Section \ref{pipline}). Conversely, in the follow-up testing phase (Section \ref{followup_result_text}), the positive class comprises patients requiring rehospitalisation, while the negative class consists of those who remained stable.

\subsection{Model Performance}
\subsubsection{Comparison of Cross-sectional and Longitudinal Modelling}
\label{ind_pair_compare}

\begin{table}[h!]
\centering

\begin{threeparttable}
\caption{Performance benchmarking (Accuracy) for the cross-sectional and LIPT scheme under varying feature selection configurations}
\label{baseline}
\setlength{\tabcolsep}{20pt}
\begin{tabular}{llllll}
\hline
\multirow{2}{*}{\textbf{\begin{tabular}[l]{@{}l@{}}Global Features\end{tabular}}} & \multirow{2}{*}{\textbf{\begin{tabular}[l]{@{}l@{}}Feature Quantity\end{tabular}}} & \multicolumn{2}{l}{\textbf{Cross-Sectional}} & \multicolumn{2}{l}{\textbf{Longitudinal Modeling}} \\
                                                                                       &                                                                                      & xgboost               & FNN                  & xgboost                  & FNN                     \\ \hline
all                                                                                  & 6373                                                                                 & 51.1                  & 55.7                 & 56.8                     & 68.2                    \\
HF-voice A                                                         & 1775                                                                                 & \textbf{69.3}         & 62.5                 & \textbf{75.0}            & 77.3                    \\
HF-voice B                                                           & 981                                                                                  & 65.9                  & \textbf{69.3}        & 70.5                    & \textbf{81.8}           \\
RASTA                                                                            & 2700                                                                                 & 48.9                  & 54.5                 & 61.4                     & 65.9                    \\
MFCC                                                                           & 1400                                                                                 & 58.0                  & 52.3                 & 61.4                     & 68.2                    \\
FFT                                                                              & 1497                                                                                 & 60.2                  & 56.8                 & 65.9                     & 68.2                    \\
ZCR                                                                            & 100                                                                                  & 53.4                  & 53.4                 & 38.6                     & 56.8                    \\
Quality                                                                         & 311                                                                                  & 52.3                  & 48.9                 & 61.4                     & 54.5                    \\ \hline
\end{tabular}

\begin{tablenotes}[flushleft]
\footnotesize
\item[a] "HF-voice A\&B" represents "HF-voice feature set A\&B" described in Sec. \ref{statistical}.
\item[b] RASTA, MFCC, FFT, ZCR and Quality are global features in this table, and 'all' means all global features combined.
\end{tablenotes}
\end{threeparttable}
\end{table}

As shown in our previous work~\cite{pan2025chinese}, the cross-sectional classification is affected by significant inter-individual variation for HF assessment using speech. In order to address this, we propose the LIPT paradigm for HF patients (defined in Section \ref{definition}). Table \ref{baseline} presents a performance comparison between cross-sectional (single-visit) and longitudinal (multi-visit) paradigms, evaluated across two distinct baseline methods: XGBoost \cite{chen2016xgboost} and Fully Connected Neural Network (FNN). To assess the discriminative power of various acoustic markers, we conducted a comparative analysis between individual global feature groups (RASTA, MFCC, FFT, ZCR and voice quality) and optimised subsets curated via rigorous statistical screening. The HF-voice feature sets A and B comprise features that attained statistical significance ($p<0.05$) in independent and paired t-tests, respectively (comprehensive statistical details are provided in Section \ref{statistical}). The results indicate that the accuracy of the proposed scheme reached 81.8\% using the FNN model, whereas the best FNN configuration for the cross-sectional scheme achieved only 69.3\%. The XGBoost method showed a similar trend, with a 5.7\% improvement observed from the cross-sectional scheme to the LIPT framework. The advantage of the proposed LIPT scheme over the cross-sectional approach aligns with the fundamental findings in our preliminary work \cite{pan2025chinese}, highlighting the effectiveness of longitudinal tracking to mitigate individual heterogeneity and achieve clinically viable monitoring. In addition, the results also demonstrate the advantage of statistically curated feature ensembles over isolated feature groups.

\subsubsection{Performance of Diverse Speech Tasks in HF Assessment}
\label{task_comparison}

\begin{figure}[]
\label{pse_task_bar}
\includegraphics[width=0.6\textwidth,trim=0 0 0 0,clip]{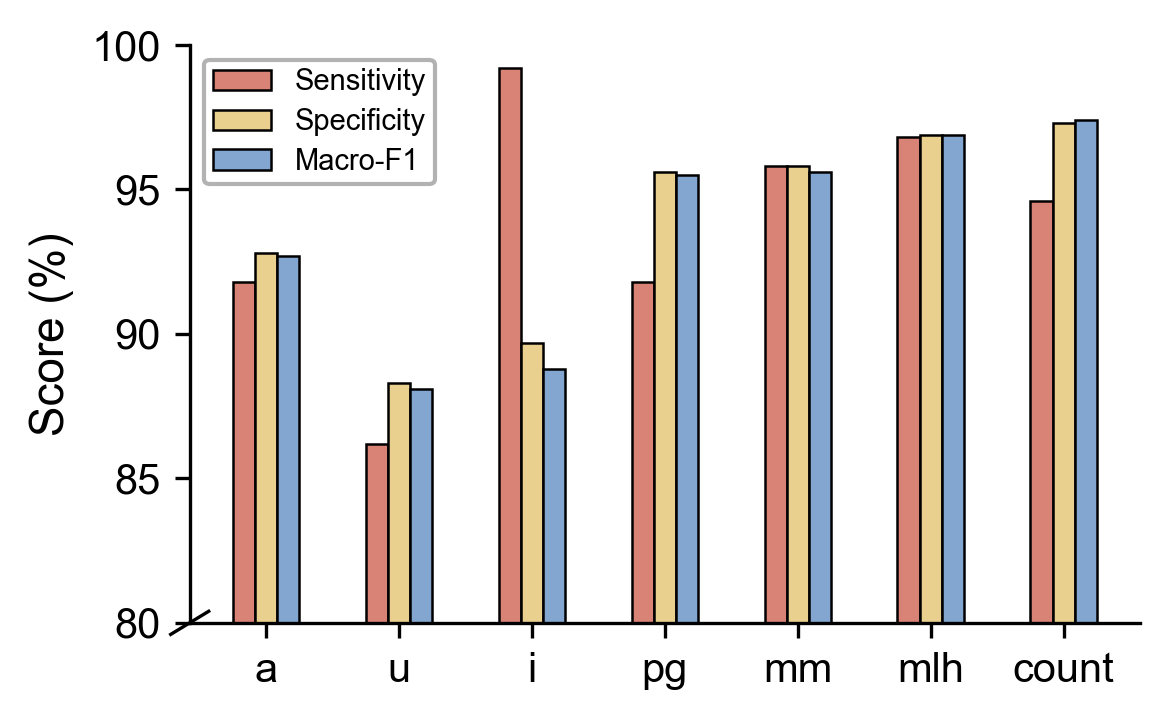}
\centering
\caption{Performance of different speech tasks with the PSE model. The long sentence task 'count' had the best macro-F1, followed by short sentences ('pg', 'mm', and 'mlh') and vowels ('a', 'i', and 'u'). The long sentence task's specificity was also the highest, albeit the sensitivity was slightly lower than that of the short sentences 'mm' and 'mlh'. These two short sentences, focusing on voiced consonants, are also relatively balanced in sensitivity and specificity.}\label{bar_rvae_task}
\end{figure}

This section evaluates the comparative effectiveness of various articulatory tasks for HF assessment. The analysis includes three vowels (sustained pronunciations of /a/, /u/, and /i/), unvoiced consonants (the short sentence 'pg'), and voiced consonants (the short sentences 'mm' and 'mlh'), while the long sentence task incorporates both voiced and unvoiced phonetic components. A comprehensive description of these speech tasks is provided in Section \ref{speech_tasks}.

The performance comparison for the PSE architecture trained on all frame-level features is detailed in Figure \ref{bar_rvae_task}. The counting task outperformed all other modalities in general, achieving a macro-F1 score of 97.4\%, which is followed closely by the three short sentences. The specificity of the counting task was also the highest, indicating a strong ability to identify the improving cases. The short sentences 'mm' and 'mlh', which focused mainly on voiced consonants, outperformed the long sentence task in terms of sensitivity with a good specificity-sensitivity balance. Although the three single-vowel tasks exhibited comparatively lower performance, they offer significant clinical advantages regarding patient compliance, as they are easier for symptomatic individuals to perform. Consequently, for the subsequent analyses in Sections \ref{pse_result} and \ref{followup_result_text}, which investigate the efficiency and generalisation properties of the PSE, evaluations are conducted using the long sentence task, as it provides the most comprehensive articulatory information.

In summary, the counting (long sentence) task is superior for intra-patient longitudinal tasks, while the performance of short sentences and vowels remains competitive. Given that the long sentence provides the most exhaustive sampling of vowels and consonants, it is evident that all types of articulatory markers captured within these tasks contribute to HF assessment. It is also worth noting that tasks based on voiced consonants also had certain advantages, especially a better sensitivity-specificity balance. In future clinical practice, speech tasks should be designed to incorporate a comprehensive set of articulations, with a particular emphasis on voiced consonants. Furthermore, a strategic balance must be maintained between achieving high diagnostic accuracy and ensuring patient compliance, especially regarding the time expenditure required for daily monitoring.

\subsubsection{Effectiveness of Personalised Sequential Encoder}
\label{pse_result}

\begin{table}[h!]
\centering
\begin{threeparttable}
\caption{Results of XGBoost, FNN, and PSE under the LIPT scheme. }
\label{rvae_result}
\setlength{\tabcolsep}{12pt} 
\begin{tabular}{lllllll}
\hline
{\textbf{Method}} & {\textbf{Features}} & Acc.          & Prec.         & Sens.         & Spec.         & F1            \\ \hline
XGBoost                          & HF-voice A                         & 75.0          & 72.0          & 81.8          & 68.2          & 74.9          \\
                                 & HF-voice B                         & 70.5          & 71.4          & 68.2          & 72.7          & 70.4          \\
FNN                              & HF-voice A                         & 77.3          & 75.0          & 81.8          & 72.7          & 77.2          \\
                                 & HF-voice B                         & 81.8          & 85.0          & 77.3          & 86.4          & 81.8          \\
PSE(ours)                        & HF-voice A                         & 64.6          & 67.1          & 70.1          & 65.6          & 64.3          \\
                                 & HF-voice B                         & 65.2          & 61.9          & 57.1          & 64.8          & 64.2          \\
                                 & all                                & 97.7          & 97.9          & 95.8          & 97.9          & 97.7          \\
                                 & MFCC                               & 97.7          & 97.8          & 96.3          & 97.8          & 97.7          \\
                                 & RASTA                              & 99.5          & 99.5          & 99.6          & 99.5          & 99.5          \\
                                 & Rhythm                             & 99.3          & 99.4          & 98.7          & 99.4          & 99.3          \\
                                 & Quality                            & 88.7          & 88.7          & 91.7          & 88.3          & 88.5          \\
                                 & RASTA+HF-voice A                   & 99.5          & \textbf{99.8} $\uparrow$ & 99.7          & 99.5          & 99.6          \\
                                 & RASTA+HF-voice B                   & \textbf{99.7} $\uparrow$ & 99.7          & \textbf{99.8} $\uparrow$ & \textbf{99.7} $\uparrow$ & \textbf{99.7} $\uparrow$ \\ \hline
\end{tabular}

\begin{tablenotes}[flushleft]
\footnotesize
\item[a] "HF-voice A\&B" represents "HF-voice feature set A\&B" described in Sec. \ref{statistical}.
\item[b] MFCC, RASTA, Rhythm and Quality are frame-level features in this table, and 'all' means all frame-level features combined.
\end{tablenotes}
\end{threeparttable}
\end{table}

To better capture temporal variations in HF status within individual patients, this work introduces a Personalised Sequential Encoder (PSE). The PSE sequentially aggregates visit-specific embeddings into a Gaussian distribution, with a reconstruction-based pre-training to initialise the encoder into a well-structured and informative spatial pattern. The structural details of the PSE are described in Section \ref{pse}. Table \ref{rvae_result} compares the performance of the proposed PSE model against the optimal feature selection (global statistical combinations) of the baseline methods described in Section \ref{baseline_method}. This comparison, conducted under the Longitudinal Intra-Patient Tracking (LIPT), focuses specifically on the transition between decompensated HF and post-treatment states. The PSE was further evaluated by incorporating the global combinations into the interim latents (see Section \ref{pse}), alongside various categories of frame-wise features. To account for the stochastic nature of the PSE model, all reported results represent the mean of five runs to ensure statistical robustness. The accuracy of the PSE model significantly exceeded that of the most effective baseline configurations. Among PSE models using a single feature category, the frame-level RASTA features proved most effective, achieving a macro-F1 score of 99.5\%. This performance was further enhanced when combined with selected global features (HF-voice feature sets A and B), yielding a sensitivity of 99.8\% and a specificity of 99.7\%. These results demonstrate high precision in detecting both clinical improvement and deterioration.

\subsubsection{Performance Validation on Follow-up} 
\label{followup_result_text}

\begin{table}[h!]
\centering
\begin{threeparttable}
\caption{Results of the follow-up test set distinguishing rehospitalised patients.} 
\label{followup_result}
\setlength{\tabcolsep}{18pt}
\begin{tabular}{llllll}
\hline
\textbf{Features} & Acc. & Prec. & Sens. & Spec. & F1 \\ \hline
all & 24.7 & 21.6 & 94.4 & 6.0 & 22.9 \\
MFCC & 25.9 & 16.4 & 61.1 & 16.4 & 25.9 \\
RASTA & 70.6 & 41.5 & 94.4 & 64.2 & 67.6 \\
Rhythm & 18.8 & 16.0 & 66.7 & 6.0 & 18.1 \\
Quality & 48.2 & 3.6 & 5.5 & 59.7 & 34.4 \\
RASTA+HF-voice A & 40.0 & 26.9 & \textbf{100} & 23.9 & 40.0 \\
RASTA+HF-voice B & \textbf{72.9} & \textbf{43.9} & \textbf{100} & \textbf{65.7} & \textbf{70.2} \\ \hline
\end{tabular}

\begin{tablenotes}[flushleft]
\footnotesize
\item[a] "HF-voice A\&B" represents "HF-voice feature set A\&B described in Sec. \ref{statistical}.
\item[b] MFCC, RASTA, Rhythm and Quality are frame-level features in this table, and 'all' means all frame-level features combined..
\end{tablenotes}
\end{threeparttable}
\end{table}

\begin{figure}[t]
     \centering
     % --- 第一行：1, 2, 3 ---
     % 使用 [t] 确保以标题首行对齐
     \begin{subfigure}[t]{0.31\textwidth}
         \centering
         \includegraphics[width=\textwidth]{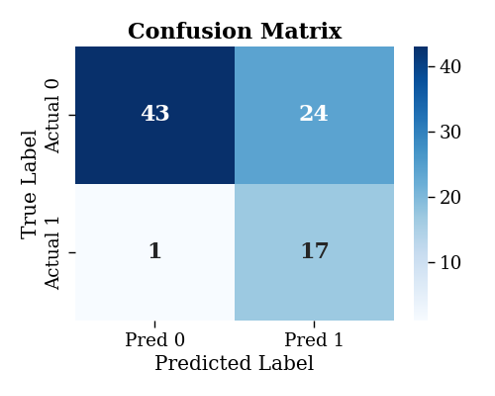}
         \caption{Confusion matrix of RASTA features.}
     \end{subfigure}
     \hfill
     \begin{subfigure}[t]{0.31\textwidth}
         \centering
         \includegraphics[width=\textwidth]{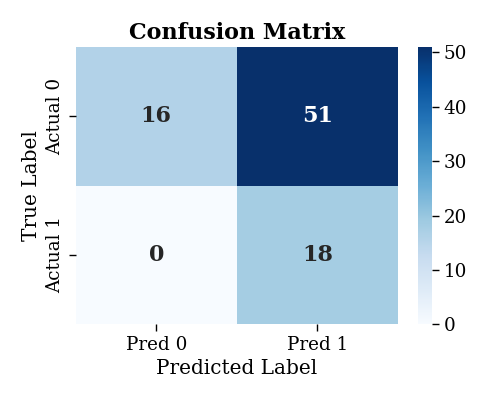}
         \caption{Confusion matrix of RASTA + Voice-HF A features.}
     \end{subfigure}
     \hfill
     \begin{subfigure}[t]{0.31\textwidth}
         \centering
         \includegraphics[width=\textwidth]{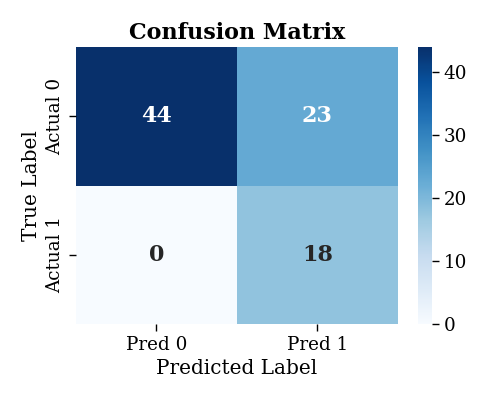}
         \caption{Confusion matrix of RASTA + Voice-HF B features.}
     \end{subfigure}

     \vspace{15pt} % 增加垂直间距，避免第一行标题撞上第二行图片

     % --- 第二行：4, 5, 6 ---
     \begin{subfigure}[t]{0.31\textwidth}
         \centering
         \includegraphics[width=\textwidth]{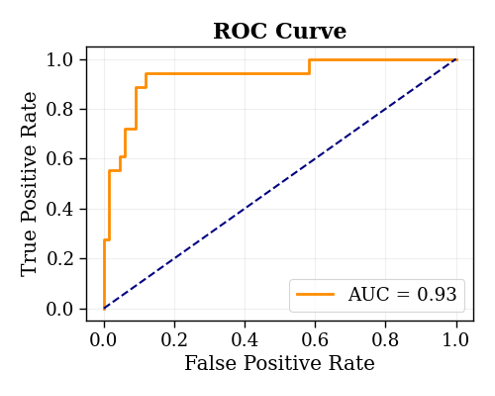}
         \caption{ROC curve of RASTA features.}
     \end{subfigure}
     \hfill
     \begin{subfigure}[t]{0.31\textwidth}
         \centering
         \includegraphics[width=\textwidth]{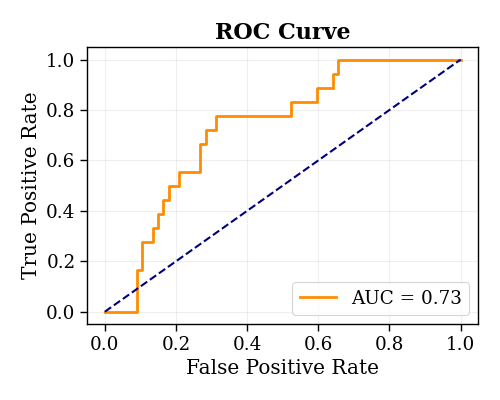}
         \caption{ROC curve of RASTA + Voice-HF A features.}
     \end{subfigure}
     \hfill
     \begin{subfigure}[t]{0.31\textwidth}
         \centering
         \includegraphics[width=\textwidth]{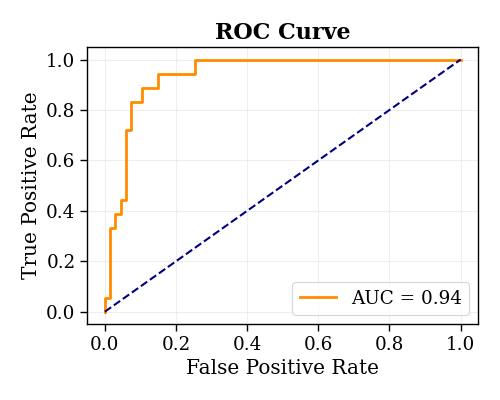}
         \caption{ROC curve of RASTA + Voice-HF B features.}
     \end{subfigure}
     
     \caption{Confusion matrix and Receiver Operating Characteristic (ROC) curve of the PSE on the follow-up data set, with the best settings overall (RASTA features with HF-voice feature sets A and B (see Section \ref{statistical})), with PSE models reported in section \ref{pse_result}. The most outstanding category of error is the false positive, indicating high confusion regarding the stable (non-hospitalised) class, where similar data is lacking in training. The relatively high area under the ROC curve (AUROC) suggests these errors stem from output misalignment rather than a lack of discriminative power.}
     \label{confusion_auroc_followup}
\end{figure}

The generalisation ability of the PSE method is evaluated by distinguishing between rehospitalised patients and those who remained stable during the follow-up phase. The models are trained on the long-sentence task between decompensated and post-treatment states, which was reported in section \ref{pse_result}, and without additional alignments for the follow-up states. For simplicity, if a rehospitalised patient also has additional records marked as stable, only the rehospitalised sample is analysed. Notably, the comparison between post-treatment and follow-up states differs from the decompensated versus post-treatment comparison in two key aspects. Firstly, in the post-treatment versus follow-up comparison, ground-truth labels are determined by the requirement for rehospitalisation (indicating deteriorating conditions), rather than being inferred from an assumed monotonic improvement over time. Secondly, the physiological and symptomatic changes between post-treatment and stable (non-rehospitalised) states are typically more subtle. This was confirmed by the NYHA records, where most patients remained at the same level. These factors jointly increase the difficulty of establishing a clear correspondence between model predictions and reference labels.

As illustrated in Table \ref{followup_result}, the RASTA-based model yielded the highest performance among the single-category configurations, achieving a macro-F1 score of 67.6\%. Performance was further enhanced by incorporating the HF-voice feature set B (see Section \ref{statistical}), with the macro-F1 score increasing to 70.2\%. Notably, the two models combining RASTA features with statistically selected global features achieved perfect sensitivity, successfully identifying all rehospitalisation cases. The confusion matrices and Receiver Operating Characteristic (ROC) curves for these three models are displayed in Figure \ref{confusion_auroc_followup}. The confusion matrices indicate that most errors were false positives, where the models incorrectly classified stable states as rehospitalised cases. A further examination of the cases in Figure \ref{confusion_auroc_followup}(a) revealed that only 3 out of the 24 false-positive cases showed improved NYHA levels, which can be viewed as 'real' errors, and the NYHA levels of the remaining 21 cases remained unchanged. Since these are not typical improvement cases as seen during training, it is expected that the model might misinterpret them as deteriorating. 

Additionally, the ROC curves suggest that the false positives may be caused by an output misalignment, as the area under the ROC curve (AUROC) remains high, reaching 0.94 in the best case. Features other than RASTA showed more significant misalignment, though they maintained reasonable sensitivities (with the exception of quality features). This indicates a strong capability in identifying rehospitalised cases, while stable cases may appear in both classes due to a lack of similar samples in the training phase. 

In summary, PSE models exhibited strong transferability in identifying rehospitalised cases during the follow-up phase. Although the model's output may shift in non-rehospitalised cases, primarily causing false positives, these errors likely stem from model misalignment on non-typical data rather than an underlying lack of discriminative ability. Beyond acquiring more typical negative data for evaluation, a more precisely defined 'stable' class is required to further align the model for non-hospitalised samples in real-world settings. 

\subsection{Discussion}
\label{discussion}
%  提出解决方法(LIPT&PSE)
Standard cross-sectional modelling methods for HF detection often suffer from limited accuracy and poor transferability due to significant inter-individual heterogeneity. In this work, we approach the HF detection problem from a longitudinal perspective, focusing on the evaluation of symptomatic changes within an individual over a specific period. Our results demonstrate that the proposed LIPT scheme achieves higher accuracy than conventional cross-sectional classification approaches, with consistent advantages observed across multiple model architectures. A specially designed PSE architecture is proposed for LIPT tasks, enabling the comprehensive encoding of a sequence of voice recordings. It achieved superior performance in longitudinal modelling compared to baseline methods and offers the distinct advantage of directly encoding individual sequential context without requiring posterior statistical information from the entire population. Furthermore, the PSE can process multiple time points simultaneously due to its sequential design, whereas baseline models are restricted to pairwise comparisons between two time points. 

% 任务设计
Although previous research has explored various speech task designs, there has been a lack of systematic discussion regarding which specific tasks are most relevant to HF detection. In comparing various articulation-based markers, minor variations in performance were observed across different speech tasks depending on the phonetic components involved. As the task with the most comprehensive set of articulations (incorporating voiced and unvoiced consonants alongside all vowels) yielded the best overall performance. The short sentence tasks involving voiced consonants also had their distinctive advantage, such as more balanced sensitivity and specificity. As a result, we recommend that new speech tasks be designed to include as many significant markers as possible, with a special emphasis on voiced consonants. Notably, even the simplest vowel-based tasks achieved accuracies of approximately 90\%, establishing them as viable alternative solutions when time is limited or patient compliance cannot be guaranteed. This finding also confirms that HF detection tasks are largely content-irrelevant, indicating that the proposed modelling methods will remain valid across different languages provided the task designs are appropriate.

% 特征选择
Another area lacking systematic guidance is the selection of acoustic features. In this work, we explored several categories proven to be associated with HF-induced articulatory changes, including time-frequency features and their derivatives, voice quality metrics, and rhythm features. We extracted both global and frame-level features, identifying distinct selection strategies for each. In our proposed PSE structure, the best performances are achieved when combining the frame-level RASTA features and the statistically selected global features, highlighting the effectiveness of both types of features. Results from both the decompensated/post-treatment phase (Table \ref{rvae_result}) and the follow-up phase (Table \ref{followup_result}) consistently identify RASTA as the most promising and relevant \textbf{frame-level} single feature group for HF monitoring, with a performance not far behind the best settings of hybrid features. Conversely, for \textbf{global} features (see Table \ref{baseline}), the results indicate that no single feature type is sufficient for optimal performance. It is even true for the global-level RASTA features, which are also statistically significant (see Figure \ref{selection}). Instead, the most effective method for constructing global HF-relevant feature sets is through building feature combinations by statistical testing (detailed in Section \ref{statistical}).

Regarding the reasons behind RASTA’s high relevance to HF detecting, apart from preserving a significant volume of spectral information, RASTA’s superior capacity for HF identification likely stems from its ability to suppress stationary noise and slow channel distortions by focusing on the modulation frequencies most critical to human respiration and articulation \cite{hermansky2002rasta}. In particular, previous work has found that RASTA features can capture pathological changes in vocal fold vibration \cite{alsulaiman2014voice}, which is also linked to HF-induced oedema. This observation is further supported by the strong performance of rhythm-related features, which rank marginally behind RASTA in Table \ref{rvae_result}. As rhythm features provide direct information regarding respiratory and speech rates, these results empirically confirm that such metrics are critical physical indicators for effective HF monitoring.

% 随访期效果体现院外监测实用价值
To evaluate the model's transferability in real-world out-of-hospital monitoring scenarios, the follow-up dataset was utilised to distinguish between rehospitalised and stable patients. As shown in Table \ref{followup_result} and Figure \ref{confusion_auroc_followup}, the model accurately identifies rehospitalised (positive) cases in the follow-up data; however, stable data tend to be inconsistently classified across different model settings. Consequently, the follow-up evaluations exhibited high sensitivity but low precision due to a high rate of false positives. The main reason for the confusion in the negative class is that the stable data are not typical improving cases as seen in the training phase, as most stable patients' NYHA levels remained the same during follow-up as when discharged. Even if the condition did improve during the stable phase, the changes remain minor. As a result, misalignments for the negative category are observed in model output, but it is highly possible that a recalibrated decision threshold could yield improved results, as the high AUROC indicates that the underlying discriminative power is strong. Nevertheless, the achieved high sensitivity is of paramount clinical importance for home-based observation, as it ensures that deteriorating patients are accurately flagged, thereby providing a crucial safety net that enables timely clinical intervention.

% 总结
To sum up, our results suggest that the proposed personalised speech modelling approach has strong potential for speech-based, out-of-hospital monitoring of HF status. It achieves high predictive accuracy while requiring only simple equipment and lightweight procedures, making it particularly attractive for settings with limited clinical capacity and constrained access to specialised care. Among the evaluated frame-level representations, RASTA emerged as the most informative feature set for HF monitoring. A plausible explanation is that RASTA emphasises perceptually meaningful spectral dynamics and suppresses slow-varying channel effects, thereby better capturing frequency regions that encode respiratory and articulatory changes relevant to HF. In contrast, for global (utterance-level) descriptors, we found that appropriate statistical feature selection strategies were critical for improving robustness and generalisability, indicating that careful dimensionality control remains essential when aggregating speech features over long time windows.

From a clinical utility perspective, the high sensitivity observed for identifying rehospitalised patients highlights the practical value of this method for early warning of deterioration and prioritisation of follow-up. Nevertheless, the current system may generate an excessive number of false-positive alerts, which could limit real-world deployability. This limitation is likely addressable through improved probability calibration, expansion to larger and more heterogeneous cohorts (particularly with more diverse follow-up intervals), and the incorporation of finer-grained clinical labels that better reflect intermediate states and symptom trajectories. Together, these findings support the feasibility of personalised speech modelling as a scalable component of remote HF monitoring, while also delineating clear directions for strengthening reliability in prospective deployments.

\section{Methods}
% 张瀚月撰写
\subsection{Data Collection}
\label{collection}

\begin{figure*}[t]
\centering
\includegraphics[width=1.05\textwidth,trim=90 0 0 0,clip]{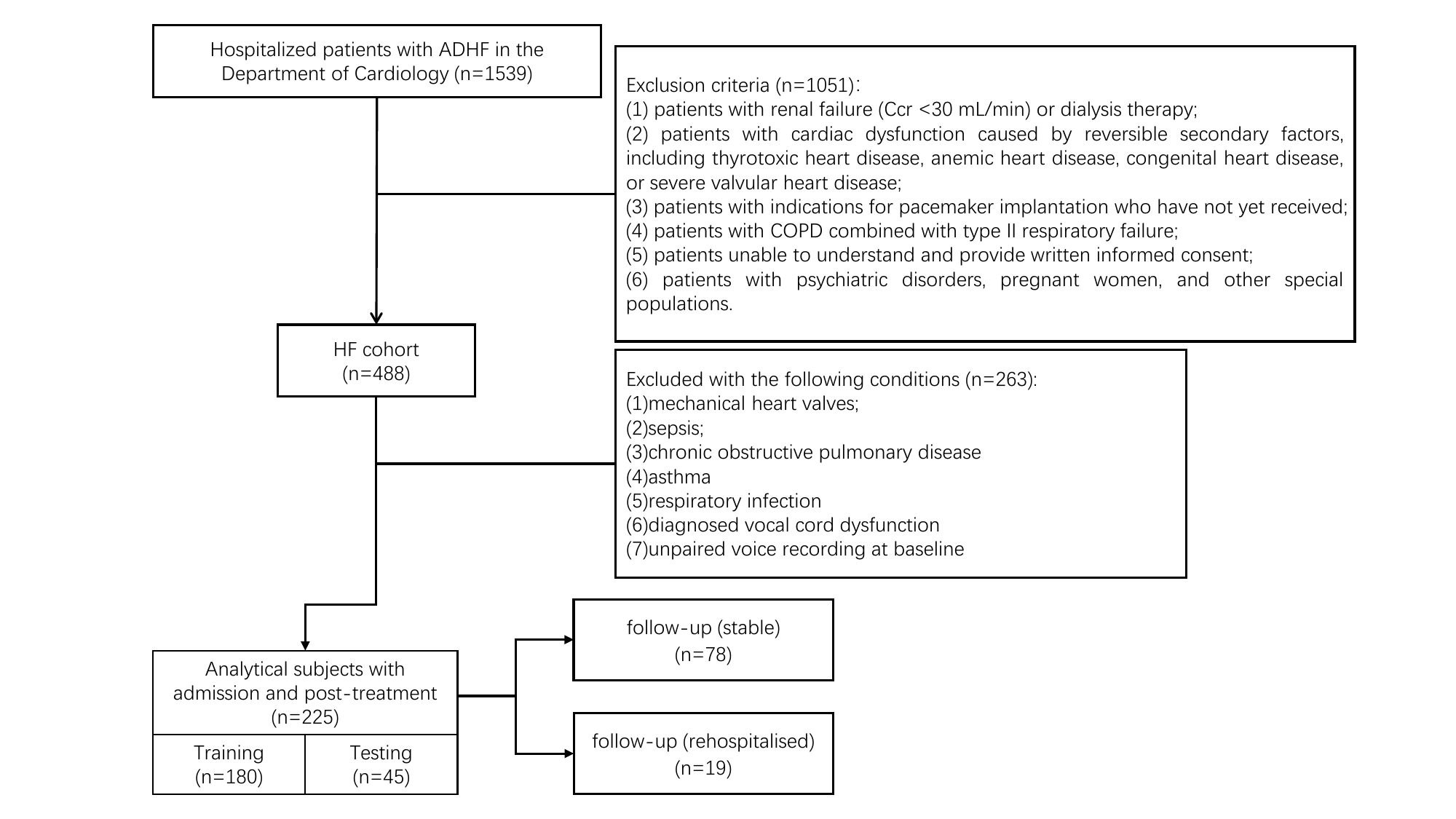}
\caption{Flowchart of the data collection process. }\label{flowchart}
\end{figure*}
%流程图225后逻辑%

\subsubsection{Study participants}
This prospective cohort study enrolled patients aged over 18 years who were hospitalised for acute decompensated heart failure (ADHF) in the Department of Cardiology, Taizhou People’s Hospital, between July 1, 2023, and October 31, 2025. Exclusion criteria were as follows: (1) patients with renal failure (Ccr <30 mL/min) or those undergoing dialysis therapy; (2) patients with cardiac dysfunction caused by various reversible secondary factors, including thyrotoxic heart disease, anemic heart disease, uncorrected congenital heart disease, or severe valvular heart disease (including severe aortic or mitral stenosis or regurgitation); (3) patients with indications for pacemaker implantation who have not yet received one; (4) patients with chronic obstructive pulmonary disease combined with type II respiratory failure; (5) patients unable to understand and provide written informed consent; and (6) patients with psychiatric disorders, pregnant women, and other special populations. During hospitalisation, patients underwent face-to-face interviews. Medical records were reviewed, and centralised data extraction was performed following a uniform data standard. Blood samples were collected for laboratory testing.

\subsubsection{Speech data recording}
\label{speech_tasks}
Electronic medical record information of patients during their hospitalisation was collected in this study, including demographic characteristics (age, sex and occupation); comorbidities (hypertension, diabetes, coronary heart disease, renal insufficiency, hyperlipidaemia, stroke and pulmonary diseases); clinical manifestations (palpitations, chest tightness, shortness of breath, loss of appetite, peripheral oedema, lung rales, etc.); laboratory tests (complete blood count, liver and kidney function, electrocardiogram, etc.); imaging examinations (cardiac ultrasound); and treatment (antiplatelet agents, statins, nitrates, etc.). Laboratory and imaging results were based on the first (i.e., earliest) recorded values from the medical records.
Voice recordings were collected at the following time points: \textbf{decompensated} phase (admission), clinical improvement \textbf{post-treatment} (discharge), and 6 and 12 months after discharge (\textbf{stable} phase). New York Heart Association (NYHA) functional class assessments were performed by cardiologists under double‑blind conditions. Voice data collected at admission and discharge were paired as baseline data. The speech tasks included: (1) Vowels: 'a', 'i', 'u'; (2) Sentence 1: ni you yi ge mei li de mei mei. (You have a beautiful sister); (3) Sentence 2: shan dong de ping guo you da you tian. (Apples from Shandong are big and sweet); (4) Sentence 3: hao yi duo mei li de mo li hua. (What a beautiful jasmine flower); (5) Number counting: Count from 1 to 60. The voice was recorded using a Xiaomi Pad 5 Pro 12.4 in a quiet room (ambient noise $<$ 50 dBA), handheld by the research staff, at a sampling rate of 22,050 Hz. Microphone-to-mouth distance was maintained at approximately 20 cm. Participants were instructed by research staff to perform speech tasks naturally.

\subsubsection{Follow-up procedure}
A follow-up team consisting of one specialist physician and three research assistants conducted regular in-person and telephone follow-ups with patients after discharge. In-person follow-ups: Scheduled at 6 and 12 months after discharge. The assessments included voice recording, echocardiography, electrocardiography, blood sample collection (for glucose, lipids, troponin, BNP/NT‑proBNP), and questionnaires. Telephone follow-ups were scheduled at 3, 9, 15, and 18 months after discharge to collect information on clinical events (cause and timing of death or rehospitalisation) and changes in medication.

\subsubsection{Data collection outcomes}
The heart failure cohort enrolled 488 patients. 234 patients were excluded based on the following conditions: mechanical heart valves, sepsis, chronic obstructive pulmonary disease, asthma, respiratory infection, or diagnosed vocal cord dysfunction. 29 were excluded due to unpaired baseline voice recordings. Ultimately, 225 paired voice datasets were analysed. Speech tasks were performed at least once by 78 of these patients during follow-up, constituting the stable‑phase voice dataset. In addition, 19 patients who were readmitted also provided at least one record in hospital. The flowchart of the whole process is shown in Figure \ref{flowchart}. As not all participants finished all designed tasks, the actual data quantity may vary. The overview of tasks and data quantity for each task are shown in Table \ref{task_definition}.

\subsection{Problem Definition: Longitudinal Intra-Patient Tracking (LIPT)}
\label{definition}
This section formally introduces the \textbf{Longitudinal Intra-Patient Tracking (LIPT)} scheme, designed to model heart failure (HF) progression by capturing personalised trajectories across monitoring sequences of arbitrary length. Unlike static classification models, LIPT is based on the principle that clinical status is best understood through the relative change between observations. Under this framework, the global clinical trajectory is mathematically derived from a series of local pairwise comparisons, meaning the fidelity of the entire longitudinal profile is fundamentally dependent on the accuracy of these local discriminative markers.

Consider a longitudinal timeline for a given patient's treatment process, denoted as $t=\{t_0, t_1, \dots, t_T\}$, where the sequence length $T$ may vary according to the duration of clinical observation. For each patient $n \in \{1, 2, \dots, N\}$, a corresponding sequence of data points $X_n=\{x_{n0}, x_{n1}, \dots, x_{nT}\}$ is collected. The LIPT framework seeks to learn a mapping function $f$ that evaluates the relative HF status between any two temporal observations $(x_{n,i}, x_{n,j})$ where $j > i$. This pairwise transformation is defined as:

\begin{equation}
f(x_{n,i}, x_{n,j}) = \hat{y}_{n,ij},
\end{equation}

where $\hat{y}_{n,ij} \in \{0, 1\}$ serves as a binary indicator of the patient's clinical trajectory (improvement or deterioration). In this work, the transformation $f$ is implemented by the proposed PSE module, where the encoded latents of the whole sequence are first produced in an integrated manner, and then the binary classifiers give the local comparison output (details in Section \ref{pse}).

While the present study primarily focuses on the precision of these local pairwise comparisons, this accurate local discrimination serves as the fundamental prerequisite for constructing a reliable global trajectory. Depending on the specific clinical requirements, different monitoring tasks can select distinct calibration objectives and prepare corresponding training labels to supervise the global reconstruction. The objectives can be established clinical gold standards, such as the New York Heart Association (NYHA) functional level and the left ventricular ejection fraction (LVEF); or early warning indicators of deterioration such as the probability of rehospitalisation $P(R_{n,j})$. Formally, we define the global objective as:

\begin{equation}
\mathcal{J} = \min \sum_{n,j} \mathcal{L} \left( \hat{y}_{n,j}, Y_{n,j}^{GS}\right),
\end{equation}

where $Y_{n,j}^{GS}$ represents the gold-standard label at time $t_j$, and $\mathcal{L}$ is a divergence measure. 

A number of options are available to achieve the global object, depending on the nature and style of the target labels. If the final target is to obtain a global ranking, classical ranking integration methods such as the Bradley-Terry model \cite{bradley1952rank} perfectly fit the requirement. Alternatively, if $Y_{n,j}^{GS}$ is a continuous regression-style target, the alignment can be achieved through a unified end-to-end training paradigm, where the temporal weights and the mapping function are treated as learnable parameters within the neural architecture. To map the discrete paired results into a continuous global trajectory, we define the following aggregation function:
\begin{equation}
\hat{y}_{n,j} = \mathcal{G}_{\phi} \left( \sum_{i=0}^{j-1} w_{ij}(\theta) \cdot f(x_{n,i}, x_{n,j}) \right)
\end{equation}

In this formulation, the trainable variables consist of the temporal weighting parameters $\theta$ and the mapping head parameters $\phi$. Specifically, the parameter $w_{ij}(\theta)$ modulates the temporal salience of historical observations, typically implemented as a learnable decay constant or an attention-based weighting mechanism. This encoded temporal information is then processed by $\mathcal{G}$, which serves to project the synthesised longitudinal evidence onto standardised clinical scales, such as NYHA functional classifications or the probabilistic risk of rehospitalisation.

In summary, while the LIPT framework provides a robust pathway for global trajectory reconstruction, the specific fitting methods required for global alignment vary across different clinical tasks. Furthermore, given the current scarcity of high-quality longitudinal datasets with sufficient sequence length for end-to-end global validation, the experimental evaluation in this study focuses primarily on the local comparison component. Nevertheless, as established in the preceding discussion, these local pairwise assessments constitute the critical foundation of the system, and once high-fidelity local discrimination is achieved, the global results can be readily and flexibly derived to suit diverse clinical requirements.

\subsection{Model Architecture}
%1. model设计
%2. loss
%3. Inference

\begin{figure*}[t]
\centering
\includegraphics[width=1.3\textwidth,trim=0 130 0 0,clip]{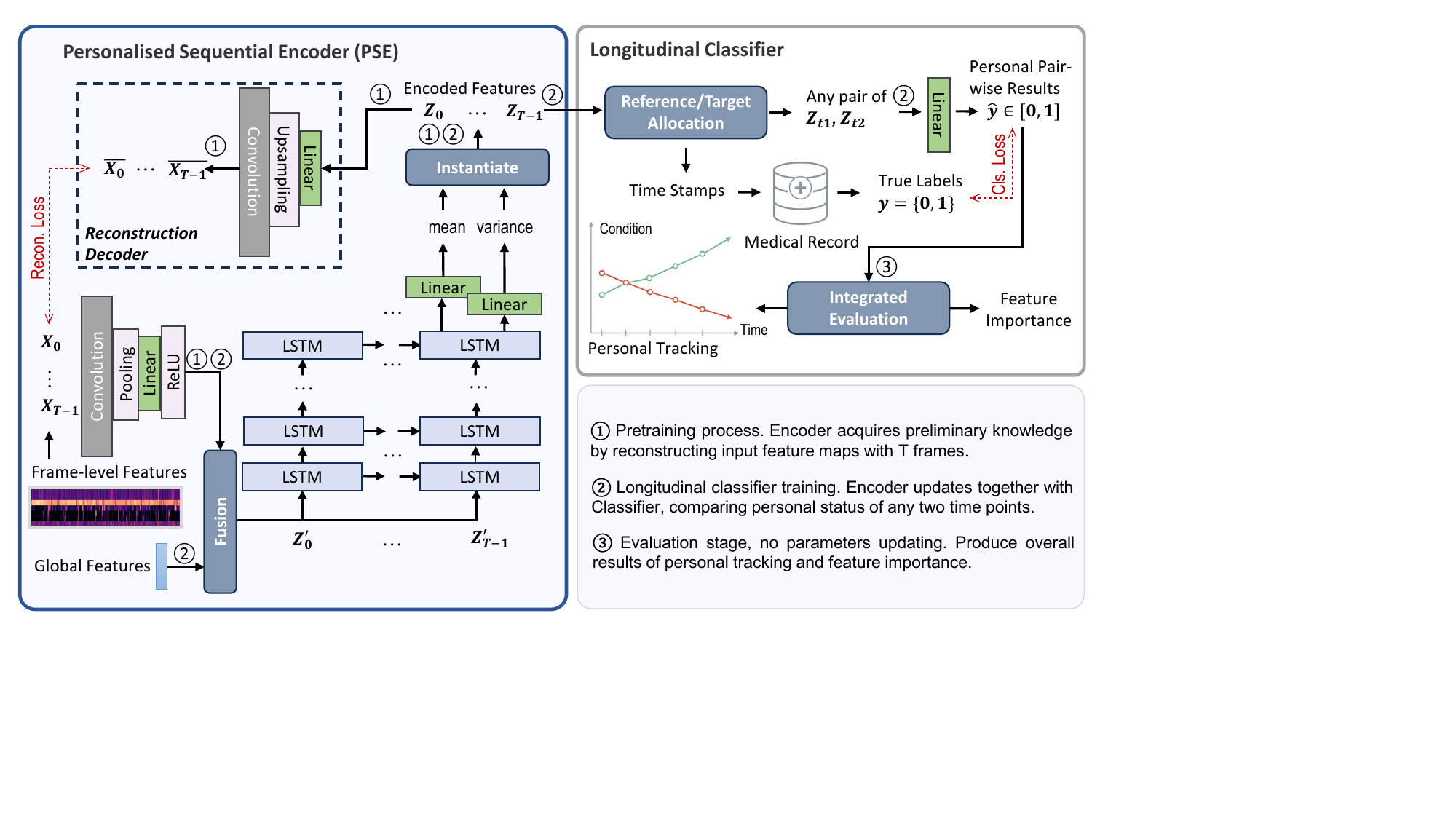}
\caption{Architecture of the Personalised Sequential Encoder (PSE) and Longitudinal Classifier for speech-based monitoring. The framework consists of three operational stages: (1) Pretraining: The encoder acquires preliminary knowledge of speech patterns by reconstructing frame-level feature maps via a decoder. (2) Longitudinal Classifier Training: Frame-level features ($X_{0} \dots X_{T-1}$) are processed through a convolutional network and aggregated with the global features to form the temporally-interpolated latent representations ($Z_{0} \dots Z_{T-1}$). The system performs reference/target allocation to compare any pair of time points ($Z_{t1}, Z_{t2}$), using a linear layer to produce local comparison results ($\hat{y} \in [0,1]$) optimised against clinical true labels. (3) Evaluation: The model integrates medical records and time stamps to produce final personal tracking results.}\label{pse_img}
\end{figure*}

\subsubsection{Baseline Models}
\label{baseline_method}
The baseline model contains three main components. Firstly, a global feature set was extracted for each recording. Then, the features were selected by their statistical significances. Finally, the XGboost \cite{chen2016xgboost} and FNN were deployed on the selected features for the classification. The HF statues can only be compared in pairs, as described in our previous work \cite{pan2025chinese}.

\subsubsection{Personalised Sequential Encoder (PSE)}
\label{pse}
To capture personalised temporal progression patterns during treatment, we designed a Personalised Sequential Encoder (PSE) (Figure \ref{pse_img}) that learns a compact and distribution-aware representation from longitudinal feature sequences. Given frame-level feature maps $X_0, X_1,...,X_{T-1}$, each time point is first mapped into a low-dimensional embedding space via a shared convolutional encoder. The global features can be used alongside the encoded features or directly used as the interim latent, skipping the convolution step entirely. These embeddings are then sequentially aggregated by a recurrent module, enabling the model to encode relative temporal dependencies and progression trends across visits. Instead of producing a deterministic latent code, the PSE models the latent representation as a probability (Gaussian) distribution ($\mathcal N(\mu,\sigma)$), which allows the encoder to account for possible variances and uncertainties in data, thus avoiding 'memorising' fixed projections instead of learning generalised latent representations. This process is done by re-parameterisation carried out by linear layers. At inference, a sample will be taken from the learnt distribution.

The encoder is pre-trained in order to retain the key information from the time-frequency domain by aligning the reconstructed feature maps with the original ones. A decoder is used to upsample the encoded feature back into the dimensions of the original features. The complete training process is described in section \ref{pipline}. During this phase, the training objective consists of a reconstruction term (MSE) that preserves informative spatial patterns, together with a distribution regularisation term (KL) that constrains the latent space to remain smooth and well-structured. The reconstruction loss function for each observation $x$ can be simplified into:
\begin{equation}
\mathcal L_{recon}
=
\underbrace{k_1(x-\mu)^2}_{\text{MSE}}
+
\underbrace{k_2\left(\mu^2+\sigma^2-\log\sigma^2-1\right)}_{\text{KL}},
\end{equation}
where $k_1$ and $k_2$ are adjustable hyperparameters. 

For the LIPT task, the learnt latent representations are directly leveraged to model temporal changes between visits. A linear predictor is applied to any given pair of latent embeddings from different time points. The estimated probability-style score is given as $\hat{y}\in[0,1]$. The target is to minimise the Binary Cross-Entropy (BCE) loss against the true label $y=\{0,1\}$: 
\begin{equation}
    \mathcal L_{\text{cls}}
=
-\big[y\log\hat y+(1-y)\log(1-\hat y)\big]
\end{equation}

In this case, the KL loss of the same form as in $\mathcal L_{recon}$ can also be added as an optional regularisation.

\subsubsection{Training Pipeline}
\label{pipline}
% Algorithm 1
This section introduces the detailed training process to obtain the longitudinal classifier. As the current training data assumes improvement over time from the decompensated state to the post-treatment state, time points are randomly arranged to create both positive and negative samples, and the models are trained to identify the original sequence. The model structure and the training process are also readily compatible with cases with multiple time points. 

Considering a time series data set $X_{raw}$, where $X_{raw_n}=\{x_{raw_{n,0}},x_{raw_{n,1}},...,x_{raw_{n,T-1}}\}$, $n\in{1,2,...,N}$ and $N$ is the total number of patients. We denote the encoder (PSE) as $\mathscr{E}$, the reconstruction decoder as $\mathscr{D}$, and the linear classifier as $\mathscr{C}$. To synchronise with the terminology used in Section \ref{definition}, the $f$ aforementioned is a combination of $\mathscr{E}$ and $\mathscr{C}$. Firstly, a feature-reconstruction pretraining is conducted to initialise $\mathscr{E}$. The feature maps $X_n$ were encoded by $\mathscr{E}$ into a global representation $Z_n$, and then decoded to reconstruct the original feature map ($\bar{X}_{raw_n}$). A combination of mean square error (MSE) loss and KL loss was calculated to update the encoder $\mathscr{E}$ and decoder $\mathscr{D}$. 

After the preliminary step, the same transformation from $X_n$ to $Z_n$ is conducted, while selected global features are merged into $Z_n$. After that, the latents of the target time points $(z_{n,i}, z_{n,j})$ are further processed by $\mathscr{C}$. A random allocation map is generated to divide encoded data into the target and reference groups, and then a subtraction is made between the target and reference samples to build the standard form training set containing both positive and negative samples. The classifier will be trained to identify the true labels, and any combination of timestamps will form a valid data pair. The parameters were updated by a binary cross-entropy (BCE) loss function. By definition, if the input pair is swapped, an opposite outcome is expected. Thus, we expand the training set by swapping each pair of target and reference data with the true label also inverted. In this way, the model will learn to be consistent regardless of the input order. The detailed steps to train the encoder, decoder, and classifier can be found in Appendix \ref{pair_data}. 

\subsection{Feature Extraction}
\label{extraction}
There are two types of features extracted: the \textbf{global} static data and the \textbf{frame-level} sequential features. The extracted feature sets are based on the ComParE 2016 \cite{schuller2016interspeech} feature sets on openSMILE \cite{eyben2010opensmile}, with additional entries, grouping, and selection. 

\subsubsection{Global Features}
A total of 6373 global features are presented and further divided into 11 groups according to internal physics (Table \ref{groups}). Group 1 (G1) contains overall statistical \textbf{spectral} features, including mean, standard deviation, skewness, kurtosis, etc. Group 2 (G2) are the same set of features, except that the signal went through Relative-Spectral \cite{hermansky2002rasta} filtering. Groups 3 (G3) and 4 (G4) are the \textbf{rhythm} features of zero-crossing rate and energy (intensity). Groups 5, 6, and 7 (G5, G6 and G7) are detailed \textbf{band-frequency} information in the form of frequency domain (by Fast Fourier Transformation), RASTA features by frequency bands, and Mel-Frequency cepstral coefficients. Group 8 contains information about the \textbf{fundamental frequency}, which may reflect both psychological \cite{yuan2002acoustic} and physical \cite{martin1995pathologic} changes in patients' voices. Finally, jitter, shimmer\cite{heiberger1982jitter}, and Harmonic-to-Noise Ratio (G9, G10, and G11) correspond to the \textbf{voice quality}, which can reflect hoarse \cite{martin1995pathologic} and breathy \cite{castillo2008automatic} voices caused by pathological changes. 

\begin{table}[]
\centering
\caption{Description of the global feature set. }\label{groups}
% \footnotesize % 保持较小字体
\setlength{\tabcolsep}{4pt} % 略微调紧列间距
\begin{tabular}{llll}
\hline
\textbf{Feature Group} &                         & \multirow{2}{*}{\textbf{Prefix in ComPare2016}} & \multirow{2}{*}{\textbf{Description}}                                                   \\ \cline{1-2}
Number                 & Name                    &                                                 &                                                                                         \\ \hline
1                      & Spectral                & audspec                                         & Spectral features                                                                       \\
2                      & RASTA                   & audspecRasta                                    & Relative-Spectral filtered \cite{hermansky2002rasta} spectral features \\
3                      & ZCR                     & pcm\_zcr                                        & Zero-crossing rate of time signal                                                       \\
4                      & RMS Energy              & pcm\_RMSenergy                                  & Root-mean-square signal frame energy                                                    \\
5                      & FFT                     & pcm\_fftMag                                     & Fast Fourier transform                                                                  \\
6                      & R-filters               & audSpec\_Rfilt                                  & RASTA features by frequency bands                                                       \\
7                      & MFCC                    & mfcc                                            & Mel-Frequency cepstral coefficients 0-14                                                \\
\multirow{2}{*}{8}     & \multirow{2}{*}{F0}     & F0final                                         & The final fundamental frequency candidate                                               \\
                       &                         & voicingFinalUnclipped                           & The voicing probability of the F0final                                                  \\
\multirow{2}{*}{9}     & \multirow{2}{*}{Jitter} & jitterLocal                                     & The local (frame-to-frame) Jitter \cite{heiberger1982jitter}           \\
                       &                         & jitterDDP                                       & The differential frame-to-frame Jitter                                                  \\
10                     & Shimmer                 & shimmerLocal                                    & The local (frame-to-frame) Shimmer \cite{heiberger1982jitter}          \\
11                     & HNR                     & logHNR                                          & Harmonic-to-noise ratio \cite{castillo2008automatic}                   \\ \hline
\end{tabular}
\end{table}

\subsubsection{Frame-level Features}
For frame-level features, there are a total of 72 features. The extractor has a window of 200$ms$, and a step length of 100$ms$. The full list of features can be retrieved in Table \ref{tab:features} in Appendix \ref{appendix_features}. These features can be categorised into 5 groups (Table \ref{groups_sequential}), including \textbf{rhythm} (energy-related and zero-crossing rate-related), \textbf{quality}, \textbf{band-frequency} features of MFCC and RASTA. Features were grouped according to their name labels in the feature set, and those with higher correlation were also grouped together. The correlation is calculated as the absolute value of the Pearson correlation between each pair of features, then averaged across all samples. Let $X = \{ X_1, X_2, \ldots, X_N \}$ be a collection of $N$ time-series data samples, where each data sample $X_i = [x_{i1}, x_{i2}, \ldots, x_{iT}]^{T}$ contains $T$ observations. For each dataset $X_i$, we compute its correlation matrix:

\begin{equation}
    R_i = |\text{corr}(X_i)|
\end{equation}

Each element of the correlation matrix $R_i$ is defined by the Pearson correlation coefficient:
\begin{equation}
r_{i,jk} = |\frac{\text{covariance}(X_{i,j}, X_{i,k})}{\sigma_{X_{i,j}} \, \sigma_{X_{i,k}}}|,
\end{equation}
where $\text{cov}(X_{i,j}, X_{i,k})$ is the covariance between variables $j$ and $k$ in dataset $i$, and 
$\sigma_{X_{i,j}}$, $\sigma_{X_{i,k}}$ are their standard deviations. Finally, the average correlation matrix across all $N$ datasets is computed as:
\begin{equation}
R_{\text{out}} = \frac{1}{N} \sum_{i=1}^{N} R_i
\end{equation} 

The average correlation matrix is shown in Figure \ref{correlation_img} of Appendix \ref{correlation_appendix}.

\begin{table}[]
\centering
\caption{Description of the frame-level feature set. }\label{groups_sequential}
% \footnotesize % 保持与上表一致的缩小字体
\setlength{\tabcolsep}{10pt} % 适当调节列间距
\begin{tabular}{lll p{6.2cm} l} % 为Description预留了较宽空间
\toprule
\multicolumn{3}{c}{\textbf{Feature Group}} & \multirow{2}{*}{\textbf{Description}} & \textbf{ID of features} \\
\cmidrule(r){1-3}
No. & & Name & & \textbf{(see Appendix \ref{correlation_appendix})} \\ 
\midrule
1 & & Rhythm (energy) & Rhythm features correlated with energy & 6, 8, 36, 37, 42, 48, 50, 68, 70 \\
2 & & Rhythm (zcr) & Rhythm features correlated with zcr & 9, 38-41, 43-47, 49, 69 \\
3 & & MFCC & Mel-Frequency cepstral coefficients 0-13 & 51-64 \\
4 & & Quality & Jitter, Shimmer \cite{heiberger1982jitter}, logHNR \cite{castillo2008automatic}, cpp \cite{fraile2014cepstral} & 2-5, 65-67 \\
5 & & RASTA & Relative Spectral Transform (RASTA) \cite{hermansky2002rasta} & 7, 10-35 \\ 
\bottomrule
\end{tabular}
\end{table}

\subsubsection{Selecting HF-related Features}
\label{statistical}
Existing feature sets are not specifically designed to capture pathological information and often contain redundant features that may impede classification performance. Consequently, statistical analysis was employed to identify the most robust and relevant features. To select the most salient global features, paired and independent t-tests were conducted to determine which speech attributes exhibited statistically significant differences between decompensated and post-treatment states. The significant features identified through these tests constitute HF-voice feature sets A and B.

The independent (two-sample) t-test is used to determine whether the means of two data groups are statistically different. It assumes the two groups are independent from each other, and they follow a normal distribution of similar variances. For the $i$ th feature, the positive and negative observations are $P=[P^1_i,P^2_i,...,P^{n}_i]$ and $N=[N^1_i,N^2_i,...,N^{n}_i]$, where n is the number of observations (patients). Let $\overline{D}=\overline{P}-\overline{N}$, where $\overline{P}$ and $\overline{N}$ are their corresponding sample means. The independent t-test can be calculated as: \cite{manfei2017differences}
\begin{equation}
    T_{1}=\frac{D}{\sqrt{\frac{2}{n}} \frac{(n-1) S_{P}^{2}-(n-1) S_{N}^{2}}{2 n-2}}
\end{equation}

When data is in the form of matched pairs, a paired t-test may be more suitable. In this case, the variance of $D$ can be estimated with the difference within each pair, thus the test value can be represented as: \cite{manfei2017differences} 
\begin{equation}
    T_{2}=\frac{D}{\sqrt{\frac{1}{n(n-1)} \sum_{j=1}^{n} (D_{j}-\bar{D})}}
\end{equation}

The null hypothesis is the mean of the two groups did not show a statistical difference. In our experiment, we deployed both types of t-tests, and selected features whose p-value is smaller than 0.05, which indicates a 95\% confidence level to reject the null hypothesis. 

\begin{figure}[t]
     \centering
     % 第一张子图
     \begin{subfigure}[b]{0.49\textwidth}
         \centering
         \includegraphics[width=\textwidth]{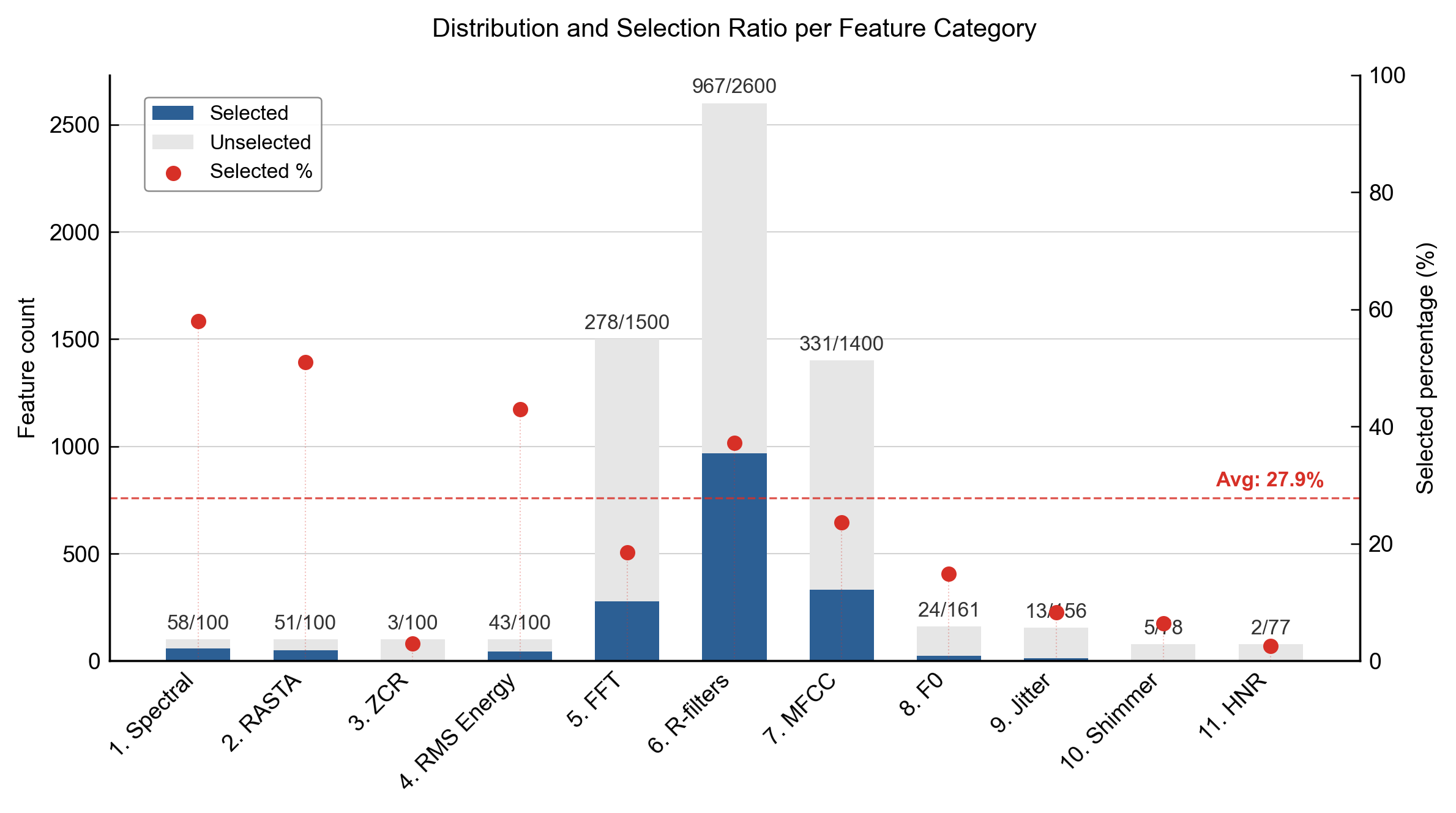}
         \caption{Significant features in the paired t-test ($p<0.05$). (HF-voice feature sets A)}
         \label{fig:left}
     \end{subfigure}
     % \hfill % 在两图之间填充弹性间距，使它们分别向左右对齐
     % 第二张子图
     \begin{subfigure}[b]{0.49\textwidth}
         \centering
         \includegraphics[width=\textwidth]{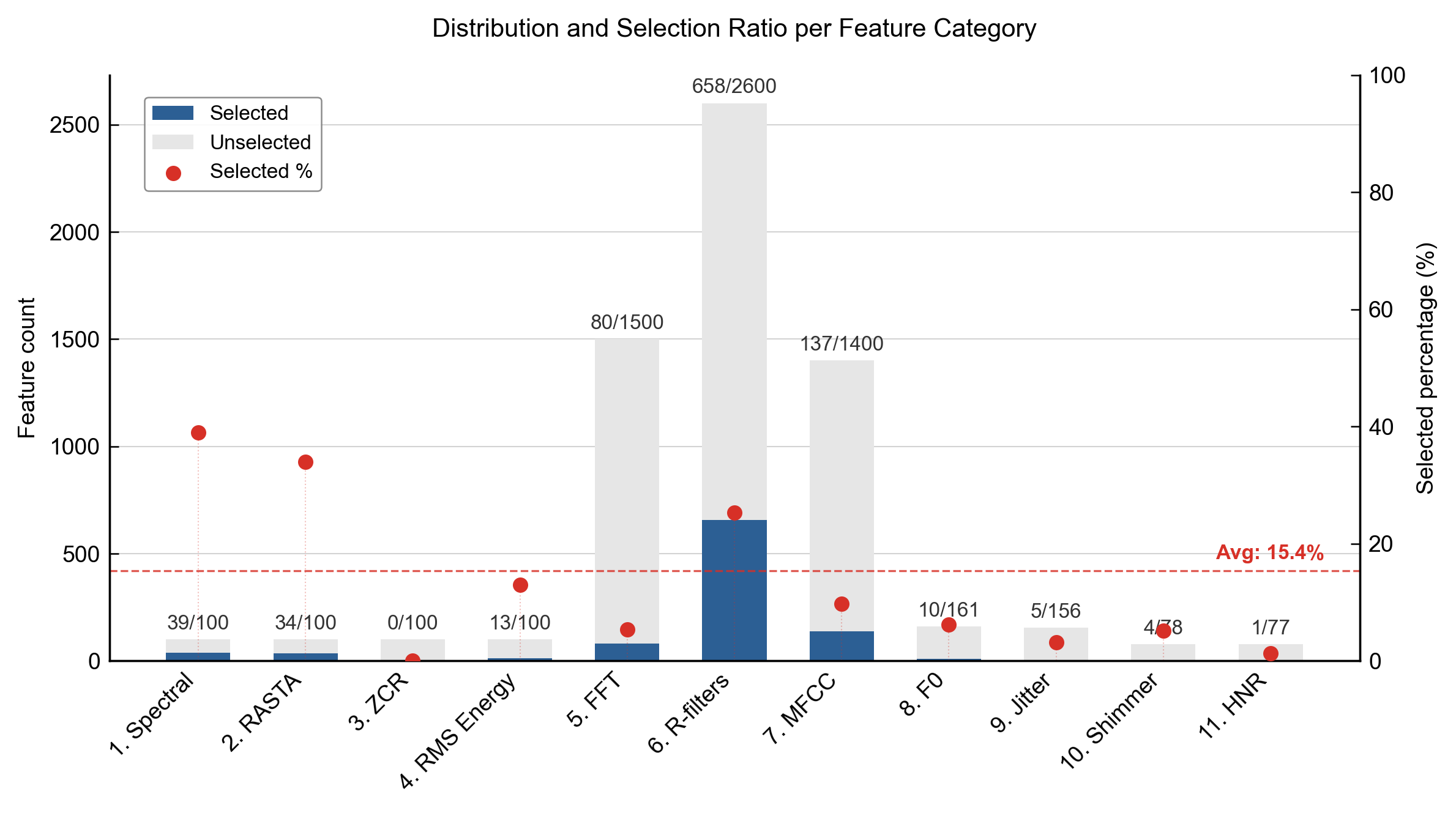}
         \caption{Significant features in the independent t-test ($p<0.05$). (HF-voice feature sets B)}
         \label{fig:right}
     \end{subfigure}
     
     \caption{Feature selection quantity and percentage with respect to the original global feature set. The spectral features (G1) are the highest in percentage in both paired and independent t-test selections, which are followed by the global RASTA features (G2). The RASTA features by frequency bands (G6) contribute the most in terms of absolute quantity in both cases.}
     \label{selection}
\end{figure}

Figure \ref{selection} illustrates the quantity and percentage of global features selected within each feature group. Among the groups defined in Table \ref{groups}, RASTA (by frequency bands) (G6) yielded the largest absolute number of features in both paired and independent t-test selections. Furthermore, spectral features (G1) related to general time-frequency information constituted the most significant category in terms of selection percentage, with global RASTA features (G2) following closely behind. This observation is consistent with the findings discussed in Section \ref{discussion}, where RASTA's importance in HF detection is corroborated by the classification results.

However, results also indicate that even those categories with high percentages of statistically significant features contain redundant information that may lead to overfitting. Consequently, a combined selection approach is necessary to improve classification efficiency for global features (Table \ref{baseline}). Finally, when conducting statistical tests, the population from which the statistics are derived must be sufficiently similar to the target population to ensure the validity and generalisability of the findings.

\subsection{Evaluation Metrics}
\label{matrix}
To comprehensively evaluate the performance and robustness of the proposed method, we employ several standard classification metrics: Accuracy (Acc.), Precision (Prec.), Recall (Sens.), Specificity (Spec.), and the Macro F1-score (F1). 

Firstly, we define the fundamental classification statistics: True Positives (TP), False Positives (FP), True Negatives (TN), and False Negatives (FN). Precision is defined as the ratio of correctly predicted positive observations to the total predicted positives. Recall, also referred to as \textbf{Sensitivity} or the \textbf{True Positive Rate (TPR)}, measures the ability of the model to identify the positive category. Specificity, also known as the \textbf{True Negative Rate (TNR)}, quantifies the model's ability to identify negative instances. These metrics and the overall accuracy are computed as follows:

\begin{align}
&\text{Precision} = \frac{TP}{TP + FP} \\
&\text{Sensitivity} = \frac{TP}{TP + FN} \\
&\text{Specificity} = \frac{TN}{TN + FP} \\
&\text{Accuracy} = \frac{TP + TN}{TP + TN + FP + FN}
\end{align}

To assess the model's classification performance while accounting for potential class imbalance, we utilise the macro F1-score. It is calculated as the harmonic mean of precision and recall, averaged over $C$ classes:

\begin{equation}
Macro-F1 =  \frac{1}{C} \sum \frac{2 \cdot P \cdot R}{P + R} 
\end{equation}

\section{Conclusion}
This study establishes a comprehensive framework for personalised heart failure monitoring using speech signals, addressing the critical challenge of inter-individual variability in human speech. By shifting the modelling paradigm from cross-sectional population analysis to longitudinal intra-patient tracking (LIPT), we demonstrate that relative vocal changes provide a more precise reflection of a patient's clinical trajectory. The proposed Personalised Sequential Encoder (PSE) successfully captures these temporal dependencies, delivering superior performance in identifying pathological changes during treatment of ADHF, as well as deterioration events during follow-up.

Our findings provide clear guidance for future research and clinical implementation regarding task design and feature selection. We recommend the inclusion of comprehensive speech tasks with an emphasis on voiced consonants to maximise diagnostic accuracy, though sustained vowels remain a viable alternative for patients with limited compliance. At the feature level, RASTA-derived frame-wise markers emerged as the most effective descriptors for capturing the subtle acoustic changes associated with HF-induced physiological alterations, and the best results are achieved when they are combined with statistically selected global features. While the current model shows high sensitivity in identifying rehospitalisation cases, future work should focus on refining the definition of 'stable' clinical states to further reduce false-positive rates. Additionally, expanding the diversity of recording environments and patient cohorts will be essential to enhance the model's generalisability across varied real-world scenarios. In summary, this personalised speech-based system offers a scalable and accurate tool for the continuous management of heart failure, particularly in resource-limited settings.

\section*{Data Availability}
Selected anonymised data is published at \url{https://github.com/panyue1998/Voice_HF_expand}. The raw data set is available upon request to our corresponding authors for research purposes, subject to approval and compliance with relevant data protection regulations. 

\section*{Code Availability}
Code and trained models are published at \url{https://github.com/panyue1998/Voice_HF_expand}.

\section*{Author contribution statement}
Y.P. and X.W. drafted the primary manuscript. H.Z. authored the data collection methodology. G.Y. and M.C. provided clinical expertise and guidance on medical practice. L.L., C.L., and R.S. oversaw data acquisition, while Y.X. and M.C. provided overall project supervision and directed the study. All authors reviewed and approved the final manuscript.

\bibliography{sample}

% \section*{Acknowledgements (not compulsory)}

% Acknowledgements should be brief, and should not include thanks to anonymous referees and editors, or effusive comments. Grant or contribution numbers may be acknowledged.

% Must include all authors, identified by initials, for example:
% A.A. conceived the experiment(s),  A.A. and B.A. conducted the experiment(s), C.A. and D.A. analysed the results.  All authors reviewed the manuscript. 
\label{MainEnd}   % ← marks last body page

\clearpage        % flush floats
\appendix

\makeatletter
  \let\OldThePage\thepage      % 备份
  \renewcommand{\thepage}{S\arabic{page}}
  \setcounter{page}{1}
\makeatother

\noindent
{\LARGE\bfseries Appendix}                
\section{Frame-level Feature List}
\label{appendix_features}
\begin{longtable}{lll}
\caption{List of frame-level features. Extracted with frame window of 200$ms$ and step length of 100$ms$.} \label{tab:features} \\
% --- 首页表头 ---
\hline
\textbf{Feature ID} & \textbf{Feature Name} & \textbf{Source} \\ \hline
\endfirsthead

% --- 续页表头 ---
\hline
\textbf{Feature ID} & \textbf{Feature Name} & \textbf{Source} \\ \hline
\endhead

% --- 续页表尾 ---
\hline
\multicolumn{3}{r}{Continued on next page...} \\
\endfoot

% --- 最后一页表尾 ---
\hline
\endlastfoot

% --- 数据开始 ---
0  & F0final\_sma & \multirow{65}{*}{\begin{tabular}[c]{@{}l@{}}Compare\_2016 \\ (openSMILE)\end{tabular}} \\
1  & voicingFinalUnclipped\_sma & \\
2  & jitterLocal\_sma & \\
3  & jitterDDP\_sma & \\
4  & shimmerLocal\_sma & \\
5  & logHNR\_sma & \\
6  & audspec\_lengthL1norm\_sma & \\
7  & audspecRasta\_lengthL1norm\_sma & \\
8  & pcm\_RMSenergy\_sma & \\
9  & pcm\_zcr\_sma & \\
10 & audSpec\_Rfilt\_sma{[}0{]} & \\
11 & audSpec\_Rfilt\_sma{[}1{]} & \\
12 & audSpec\_Rfilt\_sma{[}2{]} & \\
13 & audSpec\_Rfilt\_sma{[}3{]} & \\
14 & audSpec\_Rfilt\_sma{[}4{]} & \\
15 & audSpec\_Rfilt\_sma{[}5{]} & \\
16 & audSpec\_Rfilt\_sma{[}6{]} & \\
17 & audSpec\_Rfilt\_sma{[}7{]} & \\
18 & audSpec\_Rfilt\_sma{[}8{]} & \\
19 & audSpec\_Rfilt\_sma{[}9{]} & \\
20 & audSpec\_Rfilt\_sma{[}10{]} & \\
21 & audSpec\_Rfilt\_sma{[}11{]} & \\
22 & audSpec\_Rfilt\_sma{[}12{]} & \\
23 & audSpec\_Rfilt\_sma{[}13{]} & \\
24 & audSpec\_Rfilt\_sma{[}14{]} & \\
25 & audSpec\_Rfilt\_sma{[}15{]} & \\
26 & audSpec\_Rfilt\_sma{[}16{]} & \\
27 & audSpec\_Rfilt\_sma{[}17{]} & \\
28 & audSpec\_Rfilt\_sma{[}18{]} & \\
29 & audSpec\_Rfilt\_sma{[}19{]} & \\
30 & audSpec\_Rfilt\_sma{[}20{]} & \\
31 & audSpec\_Rfilt\_sma{[}21{]} & \\
32 & audSpec\_Rfilt\_sma{[}22{]} & \\
33 & audSpec\_Rfilt\_sma{[}23{]} & \\
34 & audSpec\_Rfilt\_sma{[}24{]} & \\
35 & audSpec\_Rfilt\_sma{[}25{]} & \\
36 & pcm\_fftMag\_fband250-650\_sma & \\
37 & pcm\_fftMag\_fband1000-4000\_sma & \\
38 & pcm\_fftMag\_spectralRollOff25.0\_sma & \\
39 & pcm\_fftMag\_spectralRollOff50.0\_sma & \\
40 & pcm\_fftMag\_spectralRollOff75.0\_sma & \\
41 & pcm\_fftMag\_spectralRollOff90.0\_sma & \\
42 & pcm\_fftMag\_spectralFlux\_sma & \\
43 & pcm\_fftMag\_spectralCentroid\_sma & \\
44 & pcm\_fftMag\_spectralEntropy\_sma & \\
45 & pcm\_fftMag\_spectralVariance\_sma & \\
46 & pcm\_fftMag\_spectralSkewness\_sma & \\
47 & pcm\_fftMag\_spectralKurtosis\_sma & \\
48 & pcm\_fftMag\_spectralSlope\_sma & \\
49 & pcm\_fftMag\_psySharpness\_sma & \\
50 & pcm\_fftMag\_spectralHarmonicity\_sma & \\
51 & mfcc\_sma{[}1{]} & \\
52 & mfcc\_sma{[}2{]} & \\
53 & mfcc\_sma{[}3{]} & \\
54 & mfcc\_sma{[}4{]} & \\
55 & mfcc\_sma{[}5{]} & \\
56 & mfcc\_sma{[}6{]} & \\
57 & mfcc\_sma{[}7{]} & \\
58 & mfcc\_sma{[}8{]} & \\
59 & mfcc\_sma{[}9{]} & \\
60 & mfcc\_sma{[}10{]} & \\
61 & mfcc\_sma{[}11{]} & \\
62 & mfcc\_sma{[}12{]} & \\
63 & mfcc\_sma{[}13{]} & \\
64 & mfcc\_sma{[}14{]} & \\ \hline
65 & cpp & \multirow{3}{*}{\begin{tabular}[c]{@{}l@{}}https://github.com/\\ satvik-dixit/CPP\end{tabular}} \\
66 & cpp\_band & \\
67 & cpp\_high & \\ \hline
68 & energy & \multirow{4}{*}{\begin{tabular}[c]{@{}l@{}}https://github.com/\\ zlzhang1124/\\ AcousticFeature\\ Extraction\end{tabular}} \\
69 & zcr & \\
70 & spl & \\
71 & activity & \\ 
\end{longtable}

\newpage
\section{Frame-level Features Correlation}

\label{correlation_appendix}
\begin{figure*}[h!]
\centering
\includegraphics[width=1.0\textwidth,trim=0 0 0 0,clip]{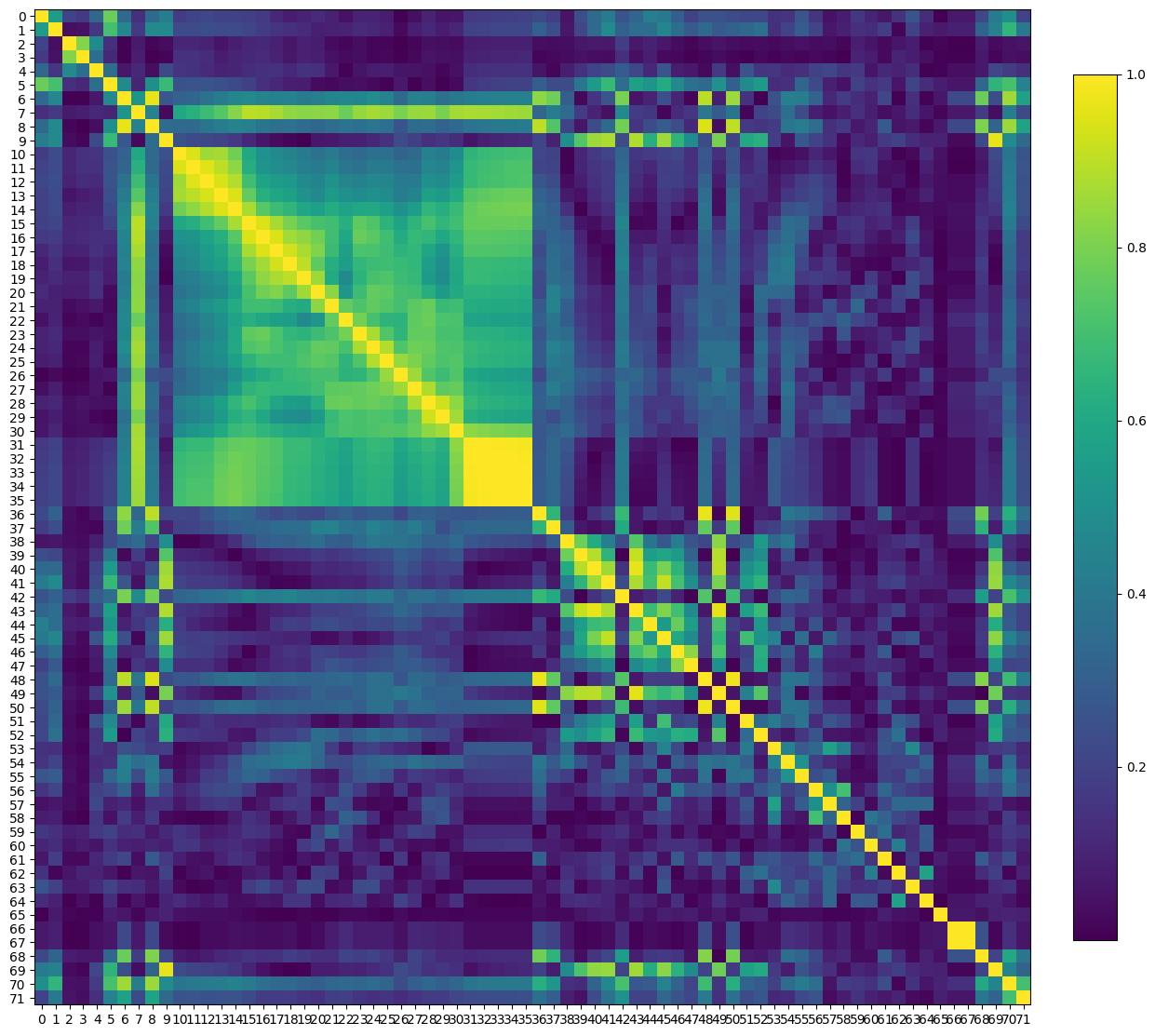}
\caption{Frame-level features' Pearson correlation matrix. Brighter colour indicate higher relevance between the two features. The numbers on each axis are the feature ID (see Appendix \ref{appendix_features}).}\label{correlation_img}
\end{figure*}

\newpage
\section{Training Algorithm of PSE}
\label{pair_data} %算法的标签
\begin{algorithm} %算法开始 
\caption{Training Pipline of the PSE model} %算法的题目 
\begin{algorithmic}
\State \textbf{Input:} $X_{raw}$
\State \textbf{Output:} $\mathscr{E}$, $\mathscr{D}$, $\mathscr{C}$
\State
\State \textbf{\#\#\# Reconstruction Pretrain}
\State $\bar{X_{raw}}=\mathscr{D}(\mathscr{E}(X_{raw}))$
\State $Loss_{recon}=MSE(\bar{X}_{raw},X_{raw})+KL(\mathcal N(\mu,\sigma^2),\mathcal N(0,1))$ 
\State \textbf{Update} $\mathscr{E}$ and $\mathscr{D}$ with $Loss_{recon}$ 
\State
\State \textbf{\#\#\# PSE forward}
\State $z_{n,0},z_{n,1},...,z_{n,T-1} = \mathscr{E}(x_{raw_{n,0}},x_{raw_{n,1}},...,x_{raw_{n,T-1}})+$selected global features 
\State 
\State \textbf{\#\#\# Classifier forward}
\State Select target time point $i,j \in(1,2,...,T)$, where $i<j$.
\State $M = [m_1,m_2,..,m_{N}], m_n =$ 0 or 1. \textit{\#create target/reference allocation map}
\State
\For{n in range $N$} \textit{\#for each patient}
    \If{$m_n==0$}
        \State $A_n = z_{n,i}$  \textit{\#reference}
        \State $B_n = z_{n,j}$  \textit{\#target}
    \EndIf
    
    \If{$m_n==1$}
        \State $A_n = z_{n,j}$ \textit{\#reference}
        \State $B_n = z_{n,i}$ \textit{\#target}
    \EndIf
\EndFor
\State
\State $X_{train} = B-A$
\State $X_{swap} = A-B$ \textit{\#for swap consistency}
\State $y_{train} = M$
\State

\State $\bar{y_{train}}=\mathscr{C}(X_{train})$
\State $\bar{y_{swap}}=\mathscr{C}(X_{swap})$ \textit{\#for swap consistency}
\State $Loss_{cls}=BCE(\bar{y_{train}},y_{train})+BCE(\bar{y_{swap}},1-y_{train})$ 
\State \textbf{Update} $\mathscr{E}$ and $\mathscr{C}$ with $Loss_{cls}$

\end{algorithmic} 
\end{algorithm}

\label{SupEnd}

\end{document}